# Probing the Role of Interlayer Coupling and Coulomb Interactions on Electronic Structure in Few-Layer MoSe$_2$ Nanostructures

Aaron J. Bradley[1,†], Miguel M. Ugeda[1,†,*], Felipe H. da Jornada[1,2,†], Diana Y. Qiu[1,2], Wei Ruan[1,3], Yi Zhang[4,5], Sebastian Wickenburg[1,2], Alexander Riss[1,‡], Jiong Lu[1,6#], Sung-Kwan Mo[4], Zahid Hussain[4], Zhi-Xun Shen[5,7], Steven G. Louie[1,2], Michael F. Crommie[1,2,8,*]

[1] Department of Physics, University of California at Berkeley, Berkeley, California 94720, USA.

[2] Materials Sciences Division, Lawrence Berkeley National Laboratory, Berkeley, California 94720, USA.

[3] State Key Laboratory of Low Dimensional Quantum Physics, Department of Physics, Tsinghua University, Beijing 100084, China.

[4] Advanced Light Source, Lawrence Berkeley National Laboratory, Berkeley, CA 94720, USA.

[5] Stanford Institute for Materials and Energy Sciences, SLAC National Accelerator Laboratory, Menlo Park, CA 94025, USA.

[6] Centre for Advanced 2D Materials and Graphene Research Centre, National University of Singapore, 6 Science Drive 2, Singapore 117546, Singapore.

[7] Geballe Laboratory for Advanced Materials, Departments of Physics and Applied Physics, Stanford University, Stanford, CA 94305, USA.

[8] Kavli Energy NanoSciences Institute, University of California Berkeley and the Lawrence Berkeley National Laboratory, Berkeley, CA 94720, USA.

* Email: mmugeda@berkeley.edu or crommie@berkeley.edu

†These authors contributed equally to this work.

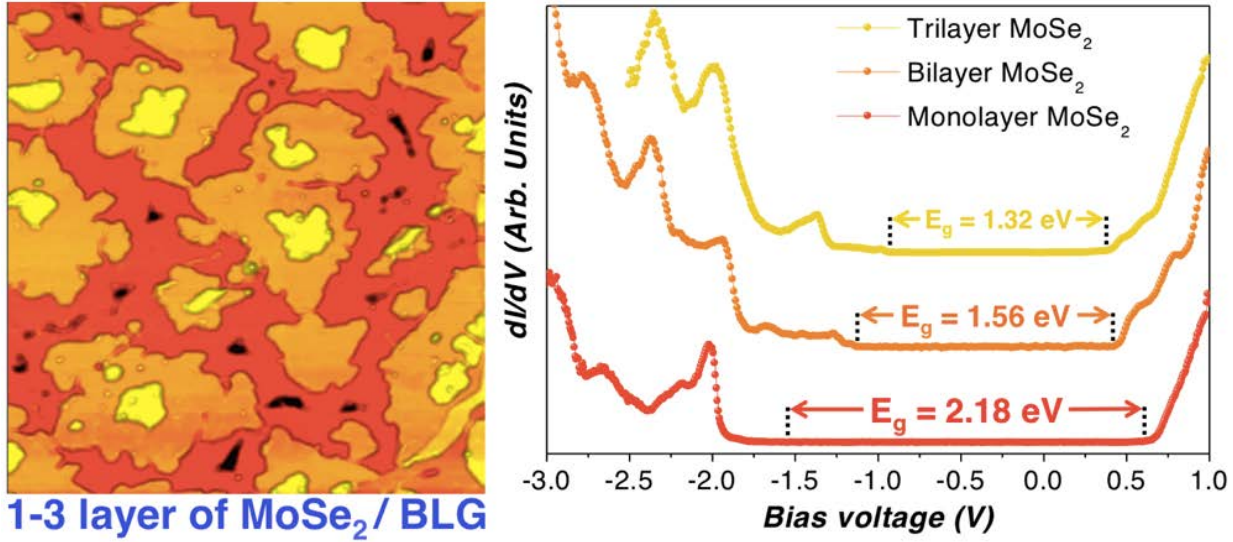

**Abstract:** Despite the weak nature of interlayer forces in transition metal dichalcogenide (TMD) materials, their properties are highly dependent on the number of layers in the few-layer two-dimensional (2D) limit. Here, we present a combined scanning tunneling microscopy/spectroscopy and GW theoretical study of the electronic structure of high quality single- and few-layer MoSe2 grown on bilayer graphene. We find that the electronic (quasiparticle) bandgap, a fundamental parameter for transport and optical phenomena, decreases by nearly one electronvolt when going from one layer to three due to interlayer coupling and screening effects. Our results paint a clear picture of the evolution of the electronic wave function hybridization in the valleys of both the valence and conduction bands as the number of layers is changed. This demonstrates the importance of layer number and electron-electron interactions on van der Waals heterostructures, and helps to clarify how their electronic properties might be tuned in future 2D nanodevices.

Owing to their inherently 2D nature, few-layer semiconducting TMDs exhibit a number of unique physical attributes that are extremely sensitive to the number of layers[1-6]. This provides new opportunities for creating van der Waals heterostructures with tailored properties and designed functionalities. For example, few-layer TMDs have been shown to support larger current densities than single layer electronic nanodevices[7], and the photovoltaic response of p-n junctions has been shown to be sensitive to the number of TMD layers[8]. Despite the promise of few-layer TMDs for electronic and optoelectronic applications, however, there is so far little understanding of how the overall electronic structure evolves with layer number close to the 2D limit. Most previous spectroscopic studies of few-layer TMD semiconductors have been via optical measurements[9-13] that only indirectly measure bandstructure, as well as photoemission[14,15] studies that primarily focus on states near the Fermi energy and in the valence band. Although many theoretical studies have been performed[16-21], a consistent picture has not yet emerged and many critical parameters, such as bandgaps and conduction band structure, remain ambiguous. In the present combined STM/STS/theory study of few-layer $MoSe_2$ on bilayer graphene, we characterize how the electronic bandgap ($E_g$), the valence band local density of states (LDOS), *and* the conduction band LDOS change with the number of $MoSe_2$ layers between 1 (monolayer (ML)), 2 (bilayer (BL)) and 3 (trilayer (TL)). These measurements compare favorably with *ab initio* GW calculations, revealing the important influence of interlayer coupling and Coulomb interactions on these properties, as well as the relative contributions from different parts of the Brillouin zone.

Low temperature (T = 5 K) STM/STS experiments were carried out on high quality $MoSe_2$ grown on bilayer graphene (BLG) on 6H-SiC(0001) substrates via molecular beam epitaxy[14]. A sketch of the structure of few-layer $MoSe_2$ is shown in Fig. 1a. Fig. 1b depicts the 2H stacking arrangement for MoSe$_2$, which we have confirmed in this study based on comparison between experiment and theory (see SI for a detailed discussion of the stacking). Samples grown with an average MoSe$_2$ coverage ranging between 0.8 and 2 ML exhibit coexisting regions of ML, BL, and TL MoSe$_2$, as well as bare BLG substrate, as shown in the STM image of Fig. 1c. Though the TL regions in Fig. 1c are relatively small compared with the ML and BL regions, we are able to tune the sizes of the different layered regions by altering the MoSe$_2$ coverage. This allowed us to maximize the area of ML, BL or TL regions and to avoid confinement and edge effects in our STS measurements.

Variations in the electronic structure between ML, BL, and TL MoSe$_2$ films on BLG were experimentally determined via STS using standard lock-in techniques[6] (all STS data were acquired at least 5 nm away from step edges, defects, and domain boundaries). Figs. 2a-c show typical STM dI/dV spectra for ML, BL, and TL MoSe$_2$, respectively. Each spectrum reveals a relatively wide bandgap surrounded by features in both the valence and conduction bands. Prominent features that determine the band edges ($V_1$, $C_1$), as well as newly discovered features in the conduction band ($C_1$, $C_2$), are marked in the spectra. Bandgap values were determined by examining the dI/dV spectra on a logarithmic scale (Figs. 2d-f) and following the statistical analysis procedure described in ref 6. For ML, BL, and TL MoSe$_2$/BLG we determine bandgap values of $E_{g,ML} = 2.18 \pm 0.04$ eV, $E_{g,BL} = 1.56 \pm 0.04$ eV, $E_{g,TL} = 1.32 \pm 0.04$ eV, respectively. Uncertainty in the values of $E_g$ arises mainly due to lateral spatial inhomogeneity and tip-induced band bending[6, 22]. Lateral inhomogeneity causes band edges to rigidly shift by 10s of meV from point to point in both the bilayer and trilayer, whereas lateral variations in ML MoSe$_2$ are smaller by an order of magnitude. These rigid shifts are presumably due to inhomogeneous doping effects and do not significantly affect the measured energy gap values.

The position of the Fermi energy ($V_{bias} = 0$ V) with respect to the band edges indicates that these samples have very low n-type doping.

In order to interpret our experimental results, we performed *ab initio* simulations of the quasiparticle electronic structure of ML, BL, and TL MoSe$_2$. These simulations allowed us to systematically study how the electronic structure of few-layer MoSe$_2$ is affected by the following factors: (1) different multilayer stacking configurations; (2) many-electron interactions; (3) interactions with the substrate; and (4) the spatial distribution of electronic states. We start by discussing the role of the stacking configuration. Five possible stacking configurations exist for two layers of MoSe$_2$. Three of these have an inversion center and two do not. We performed density functional theory (DFT) simulations for all five stacking configurations and determined that the stacking labeled AB1 (Fig. 1b) is the correct stacking sequence based on both its calculated stability and its match with experimentally observed spectroscopic features (see SI for more details). All calculations for BL and TL MoSe$_2$ were therefore performed using the AB1 structure.

Many-electron interactions were included in our calculations through the *ab initio* GW technique[23] as implemented using the BerkeleyGW package[24] (this was necessary because bare DFT does not yield accurate quasiparticle energies[23] nor optical transition energies[25]). In the first stage of the calculations we intentionally neglected the effect of the substrate by considering free-standing ML, BL, and TL MoSe$_2$. In order to speed the convergence with respect to k-point sampling, we employed nonuniform sampling of the Brillouin zone, where the smallest q-vector corresponds to ~1/1150$^{th}$ of a reciprocal lattice vector (more details in SI). To address the role of the substrate we then calculated the effect of a doped bilayer graphene substrate on supported few-layer MoSe$_2$ (the SiC was ignored because it is much less polarizable than BLG and is further away from the $MoSe_2$ layers). BLG screening of the $MoSe_2$ layers was calculated following the same method as in ref. 6.

The final calculated quasiparticle band structure (including screening contributions from the BLG substrate) for ML, BL, and TL $MoSe_2$ is plotted in the right panels of Fig. 3. This electronic structure was used to compute the LDOS above the $MoSe_2$ surface which gives a measure of the STM differential conductance (dI/dV) within the Tersoff-Hamann approximation[26] with no adjustable parameters (see SI for technical details). These theoretical STM dI/dV simulations are compared with the experimental STS spectra in the left panels of Fig. 3. We observe good agreement between the theoretical LDOS and the experimental dI/dV curves, especially near the valence band maxima (VBM) and conduction band minima (CBM). This procedure allows us to identify the reciprocal-space origin of the experimental features $V_1$, $C_1$, and $C_2$ in Fig. 2 by calculating the contributions of different regions of the Brillouin zone (see SI for more details). We are thus able to conclude that the experimental valence band feature $V_1$ originates from the K point for the ML and from the $\Gamma$ point for the BL and TL structures. The experimental conduction band features $C_1$ and $C_2$ are seen to arise from near the $\Lambda^{min}$ and $\Sigma^{min}$ points of reciprocal space, where $\Lambda^{min}$ is the point halfway between $\Gamma$ and K, and $\Sigma^{min}$ is the point halfway between $\Gamma$ and M (see inset in Fig. 3). The general good agreement between theory and experiment, especially for features close to the VBM and CBM, provides strong evidence that the main features seen in the experimental STS spectra come from the intrinsic electronic structure of $MoSe_2$ and not from extrinsic effects (such as defect states).

A comparison of our experimental and theoretical bandgaps for few-layer $MoSe_2$ can be seen in Fig. 4. The most accurate calculated bandgaps (taking into account both GW and substrate corrections) for the ML, BL, and TL structures are $2.05 \pm 0.15$ eV, $1.65 \pm 0.15$ eV, and 1.46 ± 0.15 eV respectively, within the experimental error bars. Though the magnitude of the calculated gap varies significantly with theoretical formalism, all levels of theory predict that monolayer MoSe$_2$ is a direct bandgap material at the K point of the Brillouin zone, whereas BL and TL MoSe$_2$ have indirect gaps spanning $\Gamma_v$ to $\Lambda_c^{min}$ (in contrast to some predictions that the indirect gap spans $\Gamma_v$ to K$_c$[14, 16]). These indirect transitions are also corroborated by the experimental dI/dV curves because the valence band edge signal is stronger in the BL and TL structures than in the ML structure, indicating that the valence band edge is closer to the $\Gamma$ point in BL and TL MoSe$_2$ (see V$_1$ features in Figs. 2a-c). From our calculations, the effect of the substrate reduces the direct energy gaps inhomogeneously in reciprocal space (not shown), affecting states near $\Gamma$ more than states near K. In all cases, the substrate plays a decreasing role as the number of layers is increased.

We are able to gain further insight into the electronic structure of few-layer MoSe$_2$ by examining how the spatial dependence of the simulated electronic states changes with layer number. Fig. 5 shows the modulus squared of ML and TL wave functions at the K, $\Gamma$, and $\Lambda$ points in the bandstructure. The major contribution to the conduction band state at K ($\psi_C(K)$ upper left panel) is from the Mo d orbital, so it is not expected to hybridize significantly as we add more layers. Indeed, the corresponding state $\psi_C(K)$ in the TL structure (lower left panel) looks very similar to the ML state. The valence states at K (not shown) also display little hybridization, and so we may conclude that the direct bandgap at K would not be significantly affected by hybridization between different layers. This picture is consistent with the fact that the direct gap at K predicted by DFT is constant with the number of layers (green line in Fig. 4, left panel). The picture is different for the indirect gap. The highest valence state at $\Gamma$ ($\psi_V(\Gamma)$) in ML MoSe$_2$ (upper middle panel in Fig. 5) has a significant contribution from Se p orbitals that are able to interact with similar orbitals on Se atoms in adjacent layers. Indeed, the corresponding state $\psi_V(\Gamma)$ in TL MoSe$_2$ (lower middle panel) displays significant hybridization. The lowest conduction band state at $\Lambda^{min}$ ($\psi_C(\Lambda^{min})$) is qualitatively different in the degree to which it is highly delocalized. This state is strongly modified as its spatial confinement is decreased by going from the ML to the TL structure (right panels). These differences in the character of the $\psi_V(\Gamma)$ and $\psi_C(\Lambda^{min})$ states are responsible for driving the direct-to-indirect transition in MoSe$_2$ as layer the number is increased.

In conclusion, we have measured the electronic structure of semiconducting MoSe$_2$ as a function of layer number for monolayer, bilayer, and trilayer stacking. We find that the addition of layers in the 2-dimensional regime causes the electronic bandgap to significantly shrink in size while simultaneously creating new features in both the valence and conduction bands. These experimental results are explained with theoretical GW calculations that take into account stacking geometry, wave function hybridization, electron-electron interactions, and substrate screening, thus providing new insight into how different electronic structure features arise from Bloch state properties within the Brillouin zone. The deeper understanding gained here into the electronic properties of few-layer TMD materials should help in the creation of next-generation 2D nanodevices.

**Supporting Information**. Finding bilayer stacking configuration, theoretical dI/dV calculation, electronic structure calculations. This material is available free of charge via the Internet at http://pubs.acs.org.

**Present Addresses**

‡ Institute of Applied Physics, Vienna University of Technology, Wiedner Hauptstr. 8-10/134, 1040 Wien, Austria

# Centre for Advanced 2D Materials and Graphene Research Centre, National University of Singapore, 6 Science Drive 2, Singapore 117546, Singapore

**Author Contributions**

The manuscript was written through contributions of all authors. All authors have given approval to the final version of the manuscript. The authors declare no competing financial interest.

**ACKNOWLEDGMENTS**

   This research was supported by Office of Basic Energy Sciences, Department of Energy sp$^2$ Program (STM instrumentation development and operation) and the SciDAC Program on Excited State Phenomena in Energy Materials funded by the U. S. Department of Energy, Office of Basic Energy Sciences and of Advanced Scientific Computing Research, under Contract No. DE-AC02-05CH11231 at Lawrence Berkeley National Laboratory and, which provided for algorithm and code developments and simulations. Support also provided by National Science Foundation awards no. DMR-1206512 (image analysis) and no. DMR10-1006184 (basic theory and formalism). Computational resources have been provided by the NSF through XSEDE resources at NICS and DOE at NERSC. J. L. acknowledges the National Research Foundation, Prime Minister Office, Singapore, under its Medium Sized Centre Program and CRP award "Novel 2D materials with tailored properties: beyond graphene" (R-144-000-295-281). A.R. acknowledges fellowship support by the Austrian Science Fund (FWF): J3026-N16. STM/STS data were analyzed and rendered using WSxM software[27].

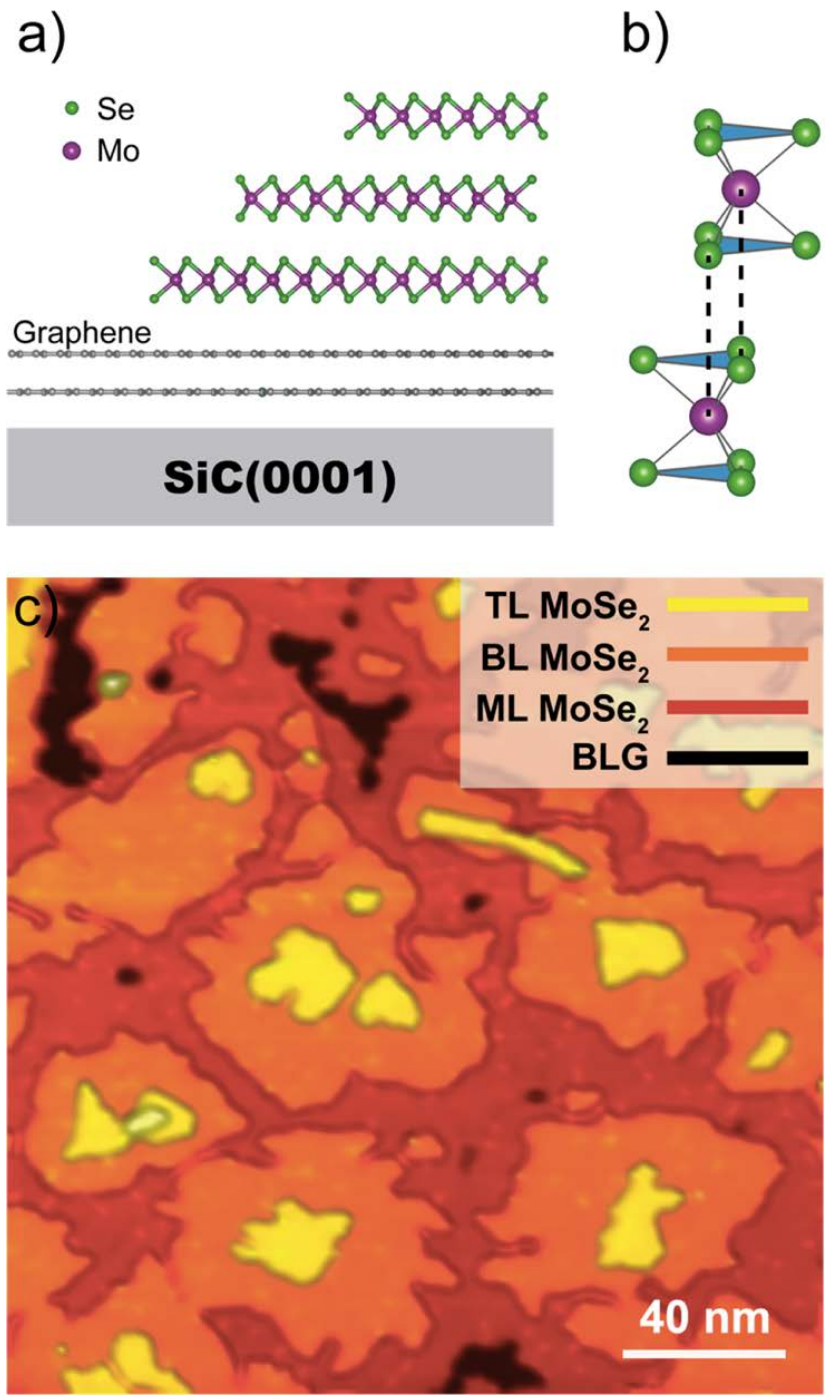

**Figure 1.** (a) Sketch of few-layer $MoSe_2$. Se atoms are shown in green, whereas Mo atoms are in purple. (b) 2H stacking configuration of $MoSe_2$ with both the Se and the Mo atoms in an AB1 stacking pattern (see SI for more details). (c) Typical STM image of 1.4 monolayer $MoSe_2$ / BLG ($V_{bias}$ = +1.5 V, $I_t$ = 30 pA, T = 5 K).

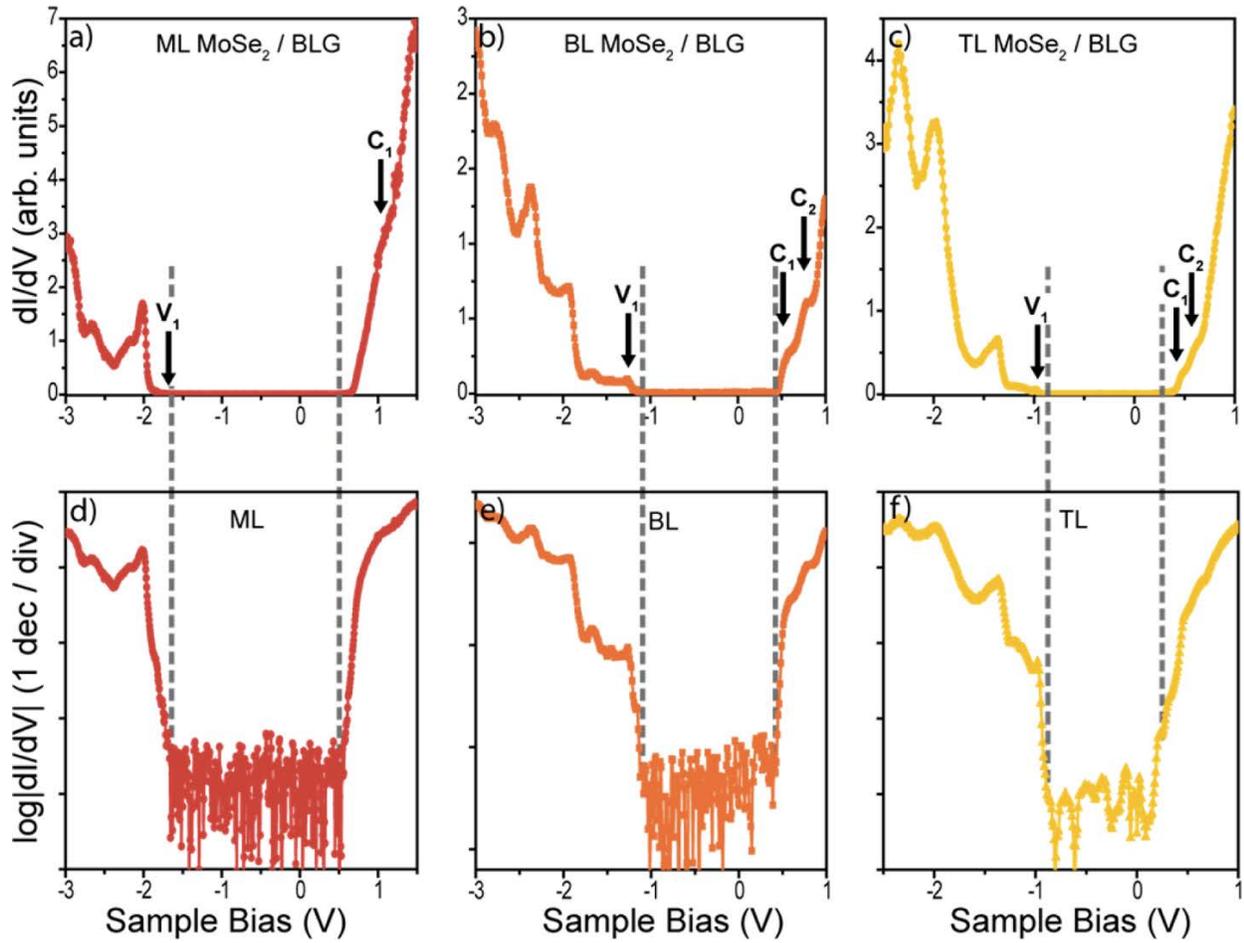

**Figure 2.** Representative STS spectra (T = 5K) obtained for (a) ML MoSe$_2$/BLG (lock-in wiggle voltage: $\Delta V_{rms}$ = 4 mV, f = 872 Hz, set point current: $I_t$ = 5 nA), (b) BL MoSe$_2$/BLG (lock-in wiggle voltage: $\Delta V_{rms}$ = 5 mV, f = 871 Hz, set point current: $I_t$ = 100 pA), and (c) TL MoSe$_2$/BLG (lock-in wiggle voltage: $\Delta V_{rms}$ = 5 mV, f = 871 Hz, set point current: $I_t$ = 5 nA). (d-f) Same STS curves shown on a logarithmic scale to highlight the electronic band edges (band edges are marked by dashed lines).

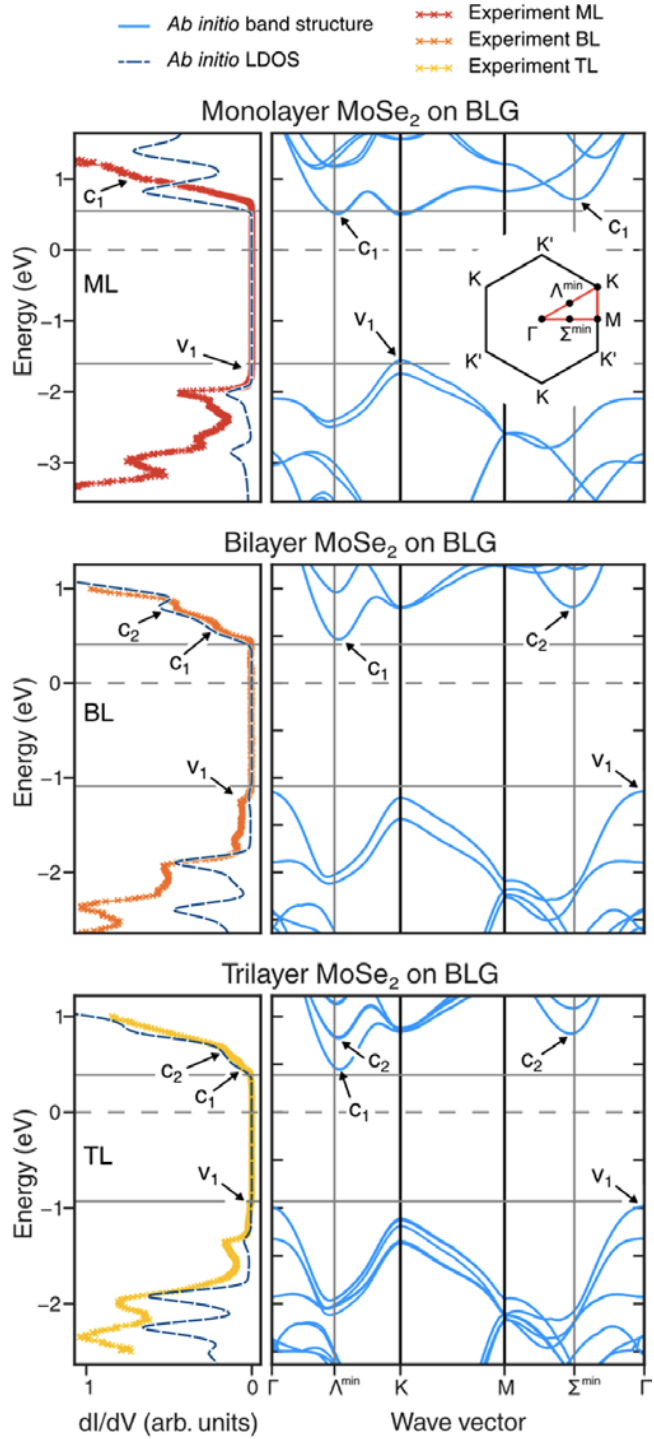

**Figure 3.** Right panels: *ab initio* GW band structure of ML, BL, and TL MoSe$_2$, including screening effects from the BLG substrate. Left panels: corresponding simulated LDOS (dashed blue lines) along with experimental STM dI/dV spectra for ML (dark red), BL (orange) and TL (yellow) MoSe$_2$. The horizontal solid lines mark the experimental VBM and CBM, and the dashed lines denote the experimental Fermi energy.

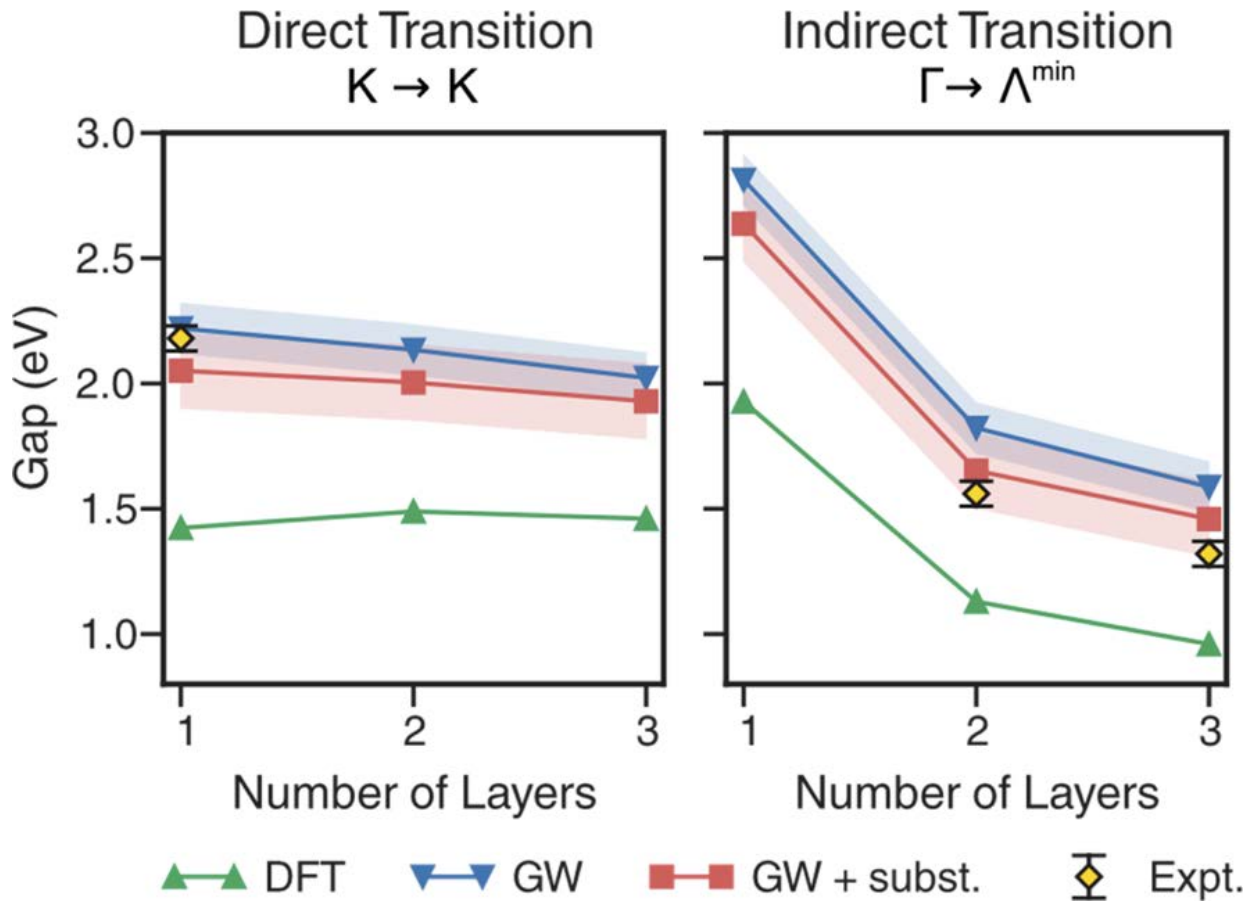

**Figure 4.** Direct and indirect bandgaps for few-layer MoSe$_2$ calculated within different levels of theory (triangles and squares) and obtained from experimental STS measurements (diamonds). The shaded regions mark the theoretical uncertainty in the GW calculations (see SI). All levels of theory predict a crossover from direct to indirect bandgap as the number of layers is increased from one to two. The theoretical uncertainty arises primarily due to the GW approximation of the electronic self-energy and the approximate treatment of the substrate (see SI).

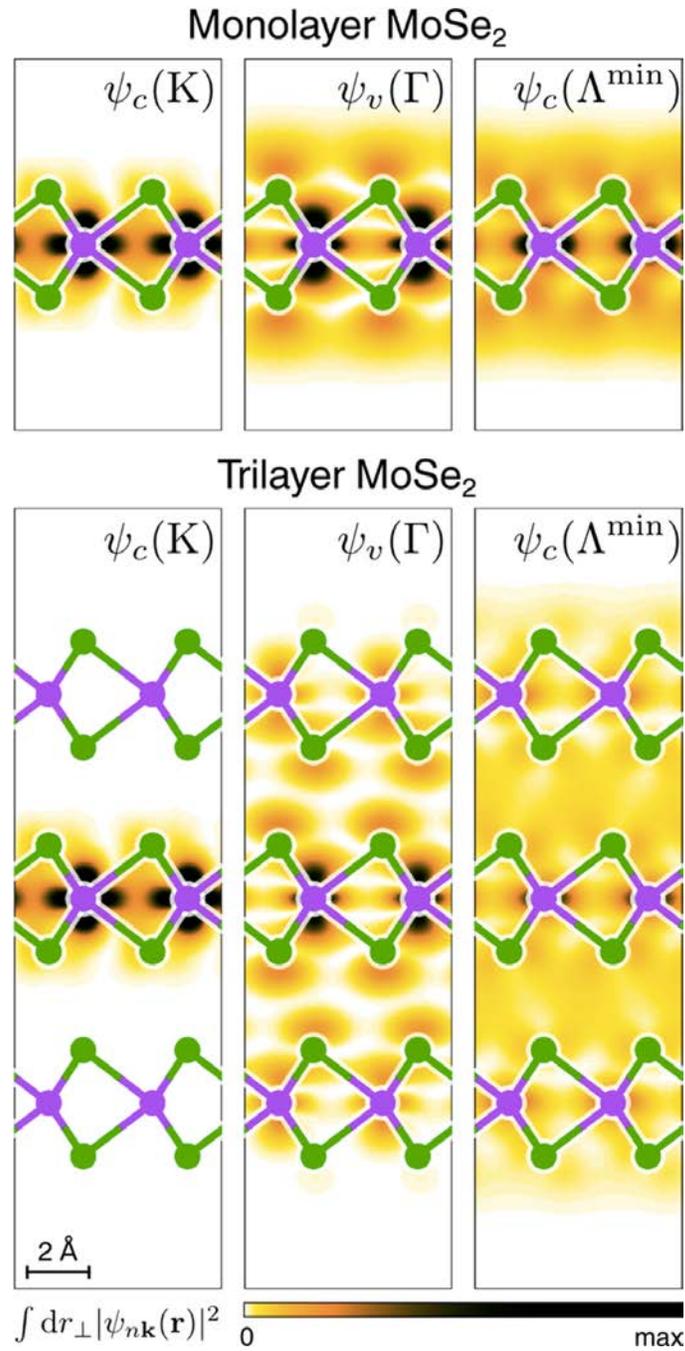

**Figure 5.** Modulus squared of calculated electronic wave functions for different Bloch states for ML and TL MoSe$_2$. The horizontal axis follows the [110] direction (same as Fig. 1a), the vertical axis points in the out-of-plane direction, and the wave functions have been integrated in the direction perpendicular to the page. Se atoms are shown in green, while Mo atoms are in purple.

Supporting Information for

# Probing the Role of Interlayer Coupling and Coulomb Interactions on Electronic Structure in Few-Layer MoSe$_2$ Nanostructures

Aaron J. Bradley[†], Miguel M. Ugeda[†,*], Felipe H. da Jornada[†], Diana Y. Qiu, Wei Ruan, Yi Zhang, Sebastian Wickenburg, Alexander Riss, Jiong Lu, Sung-Kwan Mo, Zahid Hussain, Zhi-Xun Shen, Steven G. Louie, Michael F. Crommie[*]

* Email: mmugeda@berkeley.edu or crommie@berkeley.edu

†These authors contributed equally to this work.

**Table of Contents:**

**Finding Bilayer Stacking Configuration:**

There are five possible commensurate stacking configurations for bilayer MoSe$_2$ (Figs. S1a-e). Of these, three stacking configurations, referred to as AB-stacked, have an inversion center and two configurations, referred to as AA-stacked, do not. There are two configurations (AA1 and AB1) where a Mo atom in the top layer is above a Se atom in the bottom layer. There is one configuration (AB2) where a Se atom in the top layer is above the hexagon center of the bottom layer, and two configurations (AA3 and AB3) where a Se atom in the top layer is directly above a Se atom in the bottom layer. To determine the stacking configuration of our samples, we performed fully relativistic density functional theory (DFT) calculations using the local density approximation (LDA) and scalar relativistic DFT calculations using both LDA and the van der Waals density functional with Cooper exchange (vdW-DF-c09x)[1, 2]. We find that AA1 and AB1 are the most stable stacking configurations. AA1 and AB1 also minimize the interlayer distance, suggesting that steric repulsion effects determine the stability of MoSe$_2$ stacking configurations, as is also the case in MoS$_2$[3]. The other stacking configurations, AA3, AB2, and AB3, are between 0.03 and 0.09 eV/(unit cell) higher in energy than AB1 and AA1. The Kohn-Sham band gaps and total energies of the different MoSe$_2$ stacking configurations are reported in Table 1.

Since AA1 has no inversion symmetry, spin-orbit coupling splits the valence bands near the K point into four bands that are roughly equally spaced (Fig. S1f). In the bandstructure of AB1 (Fig S1g), however, time reversal and inversion symmetries ensure that the four highest valence bands become two doubly degenerate bands, which are separated by 0.23 eV. Our STM data (Fig S2b) and previously published ARPES measurements of our bilayer MoSe$_2$ sample[4] show two valence bands at the K point separated by 0.2 eV. This strongly suggests that our sample has AB1 stacking. AB1 is also the bulk stacking configuration of 2H-MoSe$_2$. Hence, all our GW calculations are done on AB1-stacked MoSe$_2$.

**Table S1:** Total energy, Kohn-Sham band gap, valence band maximum (VBM), conduction band minimum (CBM), and bandgap at the K point for different bilayer MoSe$_2$ stacking configurations as calculated with different DFT functionals (total energy is relative to energy of AB1-stacked bilayer).

| Fully-relativistic with LDA functional | | | | | |
|---|---|---|---|---|---|
| | $E_{tot}-E_{tot}(AB1)$ (eV/unit cell) | $E_{gap}$ (eV) | VBM | CBM | $\Delta E_K$ (eV) |
| AA1 | 0.005 | 1.070 | $\Gamma$ | $\Lambda^{min}$ | 1.417 |
| AA3 | 0.082 | 1.296 | K | $\Lambda^{min}$ | 1.478 |
| AB1 | 0.0 | 1.066 | $\Gamma$ | $\Lambda^{min}$ | 1.492 |
| AB2 | 0.033 | 1.192 | $\Gamma$ | $\Lambda^{min}$ | 1.474 |
| AB3 | 0.077 | 1.327 | K | $\Lambda^{min}$ | 1.503 |
| Scalar relativistic with LDA functional | | | | | |
| | $E_{tot}-E_{tot}(AB1)$ (eV/unit cell) | $E_{gap}$ (eV) | VBM | CBM | $\Delta E_K$ (eV) |
| AA1 | 0.0 | 1.08 | $\Gamma$ | $\Lambda^{min}$ | 1.50 |
| AA3 | 0.075 | 1.44 | K | $\Lambda^{min}$ | 1.52 |
| AB1 | 0.0 | 1.07 | $\Gamma$ | $\Lambda^{min}$ | 1.54 |
| AB2 | 0.031 | 1.15 | $\Gamma$ | $\Lambda^{min}$ | 1.51 |
| AB3 | 0.075 | 1.41 | K | $\Lambda^{min}$ | 1.56 |
| Scalar relativistic with vdW-DF-c09x functional | | | | | |
| | $E_{tot}-E_{tot}(AB1)$ (eV/unit cell) | $E_{gap}$ (eV) | VBM | CBM | $\Delta E_K$ (eV) |
| AA1 | 0.0 | 1.12 | $\Gamma$ | $\Lambda^{min}$ | 1.47 |
| AA3 | 0.093 | 1.40 | K | $\Lambda^{min}$ | 1.53 |
| AB1 | 0.0 | 1.12 | $\Gamma$ | $\Lambda^{min}$ | 1.50 |
| AB2 | 0.022 | 1.20 | $\Gamma$ | $\Lambda^{min}$ | 1.51 |
| AB3 | 0.088 | 1.43 | K | $\Lambda^{min}$ | 1.53 |

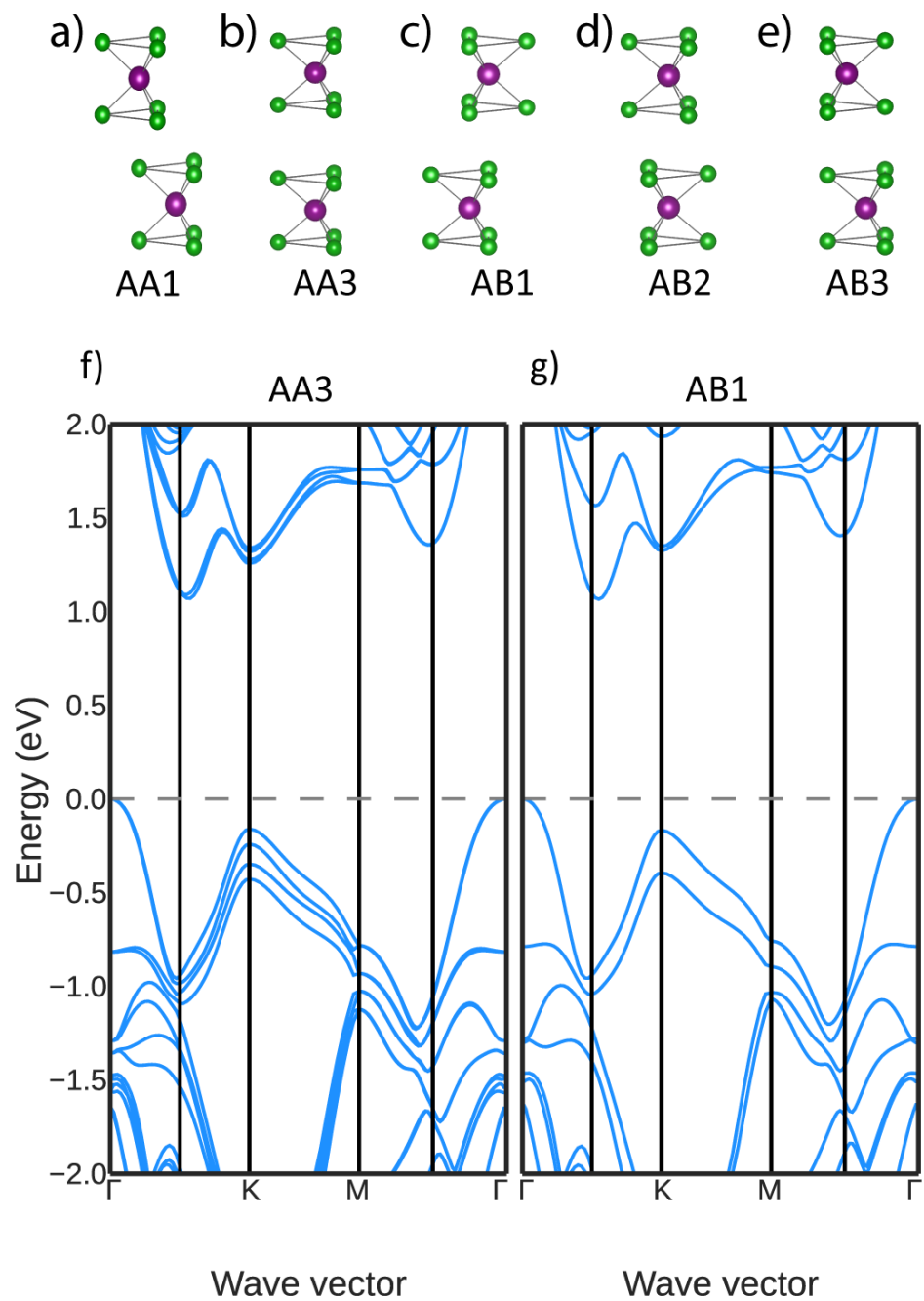

**Figure S1.** MoSe$_2$ stacking configurations: (a) AA1, (b) AA3, (c) AB1, (d) AB2, (e) AB3. (f) Bandstructure of AA1-stacked bilayer MoSe$_2$ from fully-relativistic LDA calculation. (g) Bandstructure of AB1-stacked bilayer MoSe$_2$ from fully-relativistic LDA calculation.

**Theoretical dI/dV Calculation:**

We calculated the LDOS and used it to simulate the dI/dV curve within the Tersoff-Hamann formalism[5]. For each MoSe$_2$ structure, we used a spherically localized tip 4Å above the top Se atom (center-to-center). The tip position was averaged in a plane parallel to the MoSe$_2$ surface to reduce numerical noise.

Fig. S2 shows partial contributions to the theoretical dI/dV from the $\Gamma$, K, $\Lambda^{min}$ and $\Sigma^{min}$ points in the Brillouin zone. For an MoSe$_2$ ML, this wavevector-resolved dI/dV breakdown indicates that the feature labeled V$_1$ comes from the K-point (not visible in the scale), and that the feature labeled C$_1$ comes from states from $\Lambda^{min}$ and $\Sigma^{min}$. For BL MoSe$_2$, feature V$_1$ comes from $\Gamma$, C$_1$ comes from $\Lambda^{min}$, and C$_2$ comes from $\Sigma^{min}$. For the TL, feature V$_1$ comes from $\Gamma$, C$_1$ comes from $\Lambda^{min}$, and C$_2$ comes from a combination of $\Lambda^{min}$ and $\Sigma^{min}$.

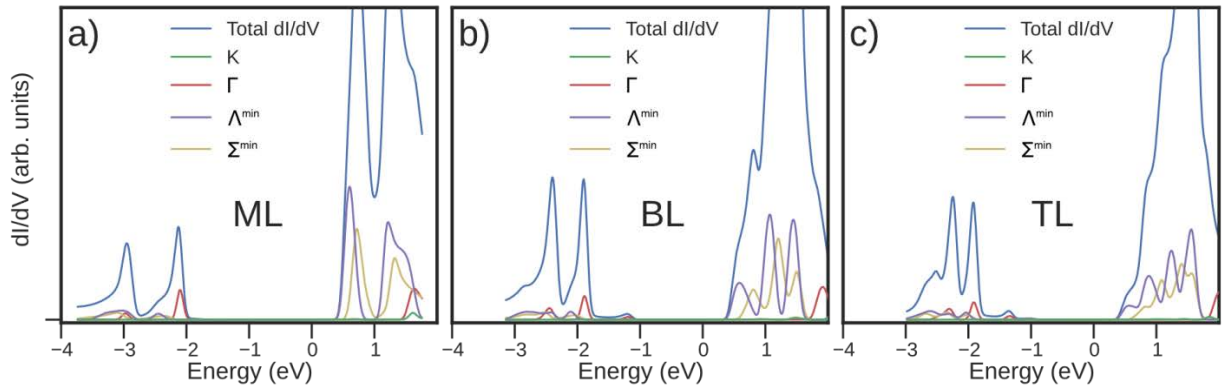

**Figure S2.** Theoretical dI/dV (blue) for the (a) ML, (b) BL, and (c) TL with partial contributions from several different points in the Brillouin zone: K (green), $\Gamma$ (red), $\Lambda^{min}$ (purple) and $\Sigma^{min}$ (yellow).

**Electronic Structure Calculations:**

We performed mean-field density functional theory (DFT) calculations in the local density approximation (LDA) using the Quantum Espresso code[6]. The calculations were done in a supercell arrangement with a plane-wave cutoff basis using norm-conserving pseudopotentials with a 125 Ry wave function cutoff. We included the Mo semicore 4d, 4p and 4s states as valence states for our DFT and GW calculations. The distance between repeated supercells in the out-of-plane direction is 50 Å. We fully relaxed the few-layer $MoSe_2$ structures including the bilayer graphene substrate, and included spin-orbit interactions as a perturbation as in Ref. (7). We calculated the substrate screening due to the bilayer graphene following the same procedure as in Ref. (8).

As mentioned in the main text, we found the GW quasiparticle bandgap to converge very slowly with respect to the number of k-points in the Brillouin zone. This slow convergence originates from the large lattice constant in the out-of-plane direction, which leads to a sharp variation in the dielectric matrix $\varepsilon_{\mathbf{GG'}}(\mathbf{q})$ for small wave vectors $\mathbf{q}$. A Monkhorst-Pack grid of at least a 60x60x1 k-points is necessary to converge the quasiparticle bandgap to within 100 meV. In order to address this computational challenge, we employed a non-uniform sampling scheme for the Brillouin zone. We first set up a regular 6x6x1 Monkhorst-Pack k-grid, and we then overlaid an extra set of 10 q-points around Γ to accurately capture the long wave length behavior of the screening for few-layer $MoSe_2$. This non-uniform approach yields the same gaps as those obtained on regular 60x60x1 k-grids, but at a small fraction of the computational cost.

We estimate our quasiparticle band structures to be converged to within 100 meV for our GW calculations. The error arises mainly from the cutoff of the dielectric matrix (20 Ry) and number of bands (5000) included in the expression for the self-energy operator. We estimate the error as roughly 150 meV for our GW calculations that include substrate screening. The extra error of ~ 50 meV is estimated from neglecting hybridization between $MoSe_2$ and BLG, as well as from the discretized frequency sampling employed to calculate the frequency-dependent dielectric response from BLG.